\documentstyle[aps,prl]{revtex}
\begin{document}
\draft
\twocolumn[\hsize\textwidth\columnwidth\hsize\csname 
@twocolumnfalse\endcsname

\title{ Hall effect of charge carriers in a correlated system}
\author{ P. Prelov\v sek }
\address{ J. Stefan Institute, University of Ljubljana, 
1001 Ljubljana, Slovenia }
\date{\today}
\maketitle
\begin{abstract}\widetext
The dynamical Hall response in a correlated electronic system is
analysed within the linear response theory for tight binding
models. At $T=0$ the d.c. Hall constant for a single quasiparticle is 
expressed explicitly via the charge stiffness, and a semiclassical 
result is recovered. As expected
a hole-like response is found for the mobile hole introduced into a
quantum antiferromagnet, as represented by the $t-J$ model.
\end{abstract}
\pacs{PACS numbers: 71.27.+a, 75.40.Cx, 71.10.+x} 
] 
\narrowtext 
The question of the Hall response in a system of correlated electrons
has proven to be extremely difficult. Theoretical investigations of
this problem have been in last decade stimulated by experiments on
superconducting cuprates, where in low-doping materials charge
carriers are holes introduced into a magnetic insulator.  Within the
normal state of cuprates this is established by Hall measurements
\cite{ong}, which reveal hole-like d.c. Hall constant $R^0_H>0$. In certain 
cases, e.g. in $La_{2-x}Sr_xCuO_4$ at low doping $x<0.15$, a simple
semiclassical result $R^0_H= 1/n_h e_0$ with hole density
$n_h=x/\Omega_0$ ($\Omega_0$ being the volume/formula unit) seems to
be obeyed at low temperatures \cite{ong}. This calls for a
semiconductor-like intepretation in terms of independent hole-like
quasiparticles (QP), rather than the usual picture for a metal with
the Fermi surface. As is however well known such descriptions fail to
explain strong $T$-dependence of $R^0_H$, persisting down to lowest
$T>T_c$ \cite{ong,carr,hwan}.

A quantum-mechanical analysis of the Hall response within the linear
response theory is complicated even for a single charge carrier, as
first analysed for the polaron problem \cite{hols} and for a single
hole in the Mott-Hubbard insulator \cite{brin}.  Analogous treatments
for metals with the Fermi surface \cite{fuku,itoh} were mainly
restricted to cases of nearly free electrons. Recently, the dynamical
Hall response $R_H(\omega)$ for the prototype models for correlated
electrons, as $t-J$ and the Hubbard model, has been approached by
analytical approximations \cite{ioff}, and in more detail evaluated by
high-$\omega,T$ analysis \cite{shas} and numerical methods
\cite{cast,assa}. Conclusions for prototype models appear 
however to be most delicate and controversial for the d.c. and
low-temperature limit $R_H^0(T\to 0)$, questioning even the hole-like
sign of $R^0_H$ in the regime of low hole doping \cite{cast,assa}.

In this Letter we treat dynamical conductivities
$\tilde \sigma_{\alpha\beta}(\omega)$ in the presence of
the magnetic field $B$, whereby we perform a linearization in $B$ 
\cite{fuku,itoh}. As first analysed by Kohn
\cite{kohn}, at $T=0$ the usual (diagonal) dynamical conductivity
(for $B=0$) is singular at low frequencies as
$\sigma_{\alpha\alpha}(\omega \to 0)\sim 2 iD_{\alpha\alpha}/\omega$,
where $D_{\alpha\alpha}$ is the charge stiffness, representing the
coherent charge response to the external field. It is now well
established that the stiffness $D$ in correlated systems is an
important and nontrivial quantity, distinguishing e.g. the metal and
the Mott-Hubbard (magnetic) insulator \cite{kohn,scal}.  We observe
that for $B>0$ also certain offdiagonal conductivities have to be
singular as well, $\sigma_{\alpha \ne
\beta}(\omega) \propto A/\omega^2$, in order to yield meaninful d.c. Hall
response $R_h^0= R_H(\omega\to 0)$.  We are able to explicitly relate
these quantitites for the case of a single mobile
carrier - QP, where we recover the simple semiclassical
relation $R^0_H=1/ne$.

Let us for simplicity consider a planar $x-y$ system, with a magnetic
field applied in the $z$ direction and with the uniform electric
current $\vec J= J_x \vec e_x$ . We follow the linear response
approach developed in Ref.\cite{fuku}, working with the field
modulated in the $y$ direction $\vec B= B {\rm e}^{iqy} \vec e_z$,
inducing a modulated electric field $\vec {\cal E} ={\cal E}_y^q {\rm
e}^{iqy} \vec e_y $.  At the final stage we are interested in the
limit $q\to 0$. As the corresponding vector potential we choose $\vec
A=A^q {\rm e}^{iqy} \vec e_x$ with $A^q= i B/q$. The dynamical Hall
response $R_H(\omega)={{\cal E}_y^q(\omega)/J_x(\omega) B}$ is given
by \cite{shas}
\begin{equation}
R_H(\omega) = {1\over B}{-\tilde \sigma^q_{yx}(\omega)\over \tilde
\sigma^q_{xx}(\omega)
\tilde \sigma^q_{yy}(\omega) -\tilde\sigma^q_{xy}(\omega)\tilde 
\sigma^q_{yx}(\omega)}\Big|_{B,q\to 0},
\label{eq1}
\end{equation}
where $\tilde \sigma_{\alpha\beta}$ denote the response at finite 
$B\ne 0$.  Models for strongly correlated electrons are usually analysed
within the tight binding framework with the magnetic field entering
(only) the kinetic energy $H_{kin}$ via the Peierls phase, i.e.
\begin{equation}
H_{kin}=- t \sum_{\langle ij\rangle s} ({\rm e}^{i \theta_{ij}}
c^{\dagger}_{js} c_{is} + H.c.), \label{eq2}
\end{equation}
where $\theta_{ij}=e\vec r_{ij}\cdot \vec A(\vec r=\vec R_{ij})$,
$\vec r_{ij}=\vec r_j - \vec r_i$ and
$\vec R_{ij}=(\vec r_i +\vec r_j)/2$. The operators for the 
particle current $\tilde
j$ and for the stress tensor $\tau$ can be now defined,
\begin{eqnarray}
\tilde j^k&=&-{1\over e}{\partial H_{kin}\over \partial 
A_{\alpha}^{-k}}= t \sum_{\langle ij\rangle s}
r^{\alpha}_{ij}{\rm e}^{i \vec k\cdot \vec R_{ij} } (i{\rm e}^{i
\theta_{ij}} c^{\dagger}_{js} c_{is} + H.c.),\nonumber \\ 
\tau^k_{\alpha\beta}&=&-{1\over e^2}{\partial^2 
H_{kin}\over \partial A_{\alpha}^{-k}\partial A_{\beta}^{-k}}= 
\nonumber \\
&=& t \sum_{\langle ij\rangle s} r^{\alpha}_{ij} r^{\beta}_{ij }
{\rm e}^{i \vec k\cdot \vec R_{ij} } 
({\rm e}^{i \theta_{ij}} c^{\dagger}_{js} c_{is} + H.c.), \label{eq3}
\end{eqnarray}
The conductivity tensor at $B\ne 0$ can be expressed as
\cite{fuku,shas},
\begin{eqnarray}
\tilde \sigma^q_{\alpha\beta}(\omega)&=&{i e^2\over \Omega
\omega^+} [\phi^q_{\alpha\beta}(0^+)  -\phi^q_{\alpha\beta}(\omega^+)],
\nonumber \\
\phi^q_{\alpha\beta}(\omega_m)&=&\int_0^\beta d\tau 
{\rm e}^{\omega_m \tau} \langle T_{\tau} \tilde j_{\alpha}^q(\tau)
\tilde j_{\beta}^0(0)\rangle_B, \label{eq4}
\end{eqnarray}
where $\omega_m=2\pi i m T$, $\omega^+=\omega+i\delta$, $\Omega$ is
the volume of the system, $T_{\tau}$ the time ordering operator and
$\langle ~\rangle_B$ denote averages for $B\ne 0$.

Next we perform the linearization in $B$, in analogy to the
treatment of the Fermi gas \cite{fuku,itoh}. From Eqs.(\ref{eq3}) it
follows
\begin{equation}
\tilde j^k_{\alpha}=j^k_{\alpha}-e \tau^{k-q}_{\alpha x}A^q,
\qquad j^k_{\alpha}= \tilde j^k_{\alpha}(B=0). \label{eq5}
\end{equation}
Taking into account the linear coupling term $H'=- e j^{-q}_x A^q$,
we can express the offdiagonal $\phi^q_{yx}$, linearized in $A^q$,
\begin{eqnarray}
\phi^q_{yx}&=& e A^q K^q_{yx}, \qquad K^q_{yx}= 
K_{yx}^I + K_{yx}^{II},\nonumber \\
K_{yx}^I(\omega_m)&=&-\int_0^{\beta} d\tau {\rm e}^{\omega_m \tau}
\langle T_{\tau} j_y^q(\tau) \tau^{-q}_{xx}(0)\rangle_0, 
\label{eq6} \\
K_{yx}^{II}(\omega_m)&=& {1\over \beta} \int_0^{\beta} d\tau 
\int_0^{\beta} d\tau' {\rm e}^{\omega_m \tau}
\langle T_{\tau} j_y^q(\tau) j_x^{-q}(\tau') j^0_x(0) \rangle_0, 
\nonumber
\end{eqnarray}
where we have taken into account that averages as  
$\langle \tau^q_{yx}(\tau)
j^0_x(0)\rangle_0$, $\langle j_y^q(\tau)j^0_x(0)\rangle_0$ 
vanish by symmetry. Within the linearization in $B$ we can 
rewrite Eq.(\ref{eq1}) as 
\begin{equation}
R_H(\omega)={e^3 (K^q_{yx}(0^+)- K^q_{yx}(\omega^+)) 
\over q ~\Omega~ \omega^+ \sigma^0_{xx}(\omega)
 \sigma^0_{yy}(\omega) }\Big|_{q\to 0}, \label{eq7}
\end{equation}
where now $\sigma_{\alpha\alpha}$ refer to the $B=0$ case.

In the following we restrict our analysis to $T=0$. Let us assume for
simplicity that the absolute ground state $|0\rangle$, with the energy
$E_0$ and corresponding to the wavevector $\vec Q=0$, is
nondegenerate. For diagonal $\sigma^q_{\alpha\alpha}$ we can perform
the $q \to 0$ limit. Strictly at $q=0$ we take into account the sum
rule $\phi^q_{\alpha\alpha}(0) \to \langle
\tau^0_{\alpha\alpha} \rangle$, while $\phi^0_{\alpha\alpha}(\omega \to
0) <\phi^q_{\alpha\alpha}(0^+)$.  One can separate the response into
the coherent (singular) part
\cite{kohn,scal} and the  incoherent (regular) part, expressed 
in term of eigenstates, 
\begin{eqnarray}  
\sigma^0_{\alpha\alpha}(\omega)&=& {2 i  e^2 \over \Omega \omega^+} 
D_{\alpha\alpha} + \sigma_{\alpha\alpha}^{reg}(\omega), \nonumber \\
D_{\alpha\alpha}&=& {\langle \tau^0_{\alpha\alpha}\rangle \over 2} -
\sum_{m>0} {|(j^0_{\alpha})_{0m}|^2 \over \epsilon_m}, \label{eq8} \\
\sigma_{\alpha\alpha}^{reg}(\omega)&=& {i e^2 \over \Omega}
\sum_{m>0} {|(j_{\alpha}^0)_{0m}| ^2 \over \epsilon_m}
[{1\over \omega^+ + \epsilon_m}  +{1\over \omega^+ - \epsilon_m}] 
\nonumber, 
\end{eqnarray}
where we use the notation $(j^0_{\alpha})_{0m}=\langle
0|j^0_{\alpha}|m\rangle$ and $\epsilon_m=E_m-E_0$.

Expressing $K^q_{yx}$ in terms of eigenstates (at $T=0$) is also
straightforward, although more tedious. Due to the $q$-character of
operators entering Eqs.(\ref{eq6}), it is convenient to distinguish
eigenstates $|m \rangle, |\tilde m \rangle$ and $|\hat m\rangle$,
corresponding to wavevectors $\vec Q,~\vec Q-\vec q$ and $\vec Q +
\vec q$, respectively. So we obtain from Eqs.(\ref{eq6}),
\begin{eqnarray}
K^q_{yx}(\omega^+)&=&\sum_{\tilde m}\Bigl[{\gamma_{\tilde m} \over
\omega^+ - \epsilon_{\tilde m}} +{\tilde \gamma_{\tilde m} \over 
\omega^+ + \epsilon_{\tilde m}} \Bigr] + \nonumber \\
&+& \sum_{m>0}\Bigl[{\delta_m \over \omega^+ - 
\epsilon_m} +{\tilde \delta_m \over \omega^+ + \epsilon_m} \Bigr], 
\label{eq9}
\end{eqnarray}
where it follows from $\tilde \sigma_{\alpha\beta}(-\omega)=
\tilde \sigma^*_{\alpha\beta}(\omega)$ and 
Eqs.(\ref{eq4},\ref{eq6},\ref{eq9})
that $\tilde \gamma_{\tilde m}=\gamma_{\tilde m} = {\rm real},~~
\tilde \delta_m = \delta_m = {\rm real}$, and
\begin{eqnarray}
\gamma_{\tilde m}&=&(j^q_y)_{0\tilde m}\Bigl[
(\tau^{-q}_{xx})_{\tilde m 0}- \sum_l {(j^{-q}_x)_{\tilde m
l}(j^0_x)_{l0} \over \epsilon_l - \epsilon_{\tilde m} }
- \nonumber \\
&&\qquad - \sum_{\tilde l} 
{(j^{-q}_x)_{\tilde l 0}(j^0_x)_{\tilde m \tilde l}
\over \epsilon_{\tilde l} } \Bigr],  \label{eq10} \\
\delta_m&=&-(j^0_x)_{m 0}\Bigl[ \sum_{\tilde l} 
{(j^q_y)_{0\tilde l} (j^{-q}_x)_{\tilde l m} \over 
\epsilon_{\tilde l} - \epsilon_m }+\sum_{\hat l} 
{(j^q_y)_{\hat l m} (j^{-q}_x)_{0 \hat l} \over
\epsilon_{\hat l}} \Bigr], \nonumber
\end{eqnarray}

We again separate $K^q_{yx}(\omega)$ into the regular and the singular
part. The latter should cancel in Eq.(\ref{eq7}) the singular terms
$\sigma^0_{\alpha\alpha} \propto 1/\omega$, in order to yield a
meaningful $R_H(\omega \to 0)$. It is evident that the relevant
contribution to $K^q_{yx}$ comes in Eq.(\ref{eq9}) from terms with
$\epsilon_m, \epsilon_{\tilde m} \rightarrow 0$. In analogy to free
fermions we can speculate on several possibilities for low lying
excitations. E.g. in a metal with the Fermi surface one has to
consider electron-hole pairs as the relevant excited states. At
present we are not able to treat Eq.(\ref{eq9}) in a meaningful way
for an analogous regime in a correlated metal.

It is however feasible to consider the nontrivial response of a single
charge carrier - QP, e.g. introduced by doping the Mott-Hubbard
insulator or antiferromagnet (AFM). For a well defined QP we require a
quadratic dispersion $\epsilon_{q \to 0} \propto q^2$ and a pseudogap
in the optical response $\sigma^{reg}_{\alpha\alpha}(\omega \to 0)\to
0$, hence $(j^0_{\alpha})_{0m}\to 0$ for $\epsilon_m \to 0$.  Such
assumptions seem to hold e.g. for a mobile hole introduced into a 2D
quantum AFM \cite{schm,prel,eder}.

Under these restrictions we note that $\delta_m$ terms contribute only
to $(\tilde \sigma^q_{yx})^{reg}$, due to the prefactor
$(j^0_{\alpha})_{0m}$, which vanishes for $m \to 0$. In contrast, the
essential contribution to the ${\tilde m}$ sum in Eq.(\ref{eq9}) comes
from the QP excited state $|\tilde 0 \rangle$ with $\epsilon_{\tilde
0} =\epsilon_q \agt 0$. It is possible to simplify $\gamma_{\tilde
0}$, since we can at the same time perform the limit $q \to 0$ for
certain matrix elements (note that $\vec q$ points in the
$y$-direction), i.e.  $(\tau^{-q}_{xx})_{\tilde 00} \to
(\tau^0_{xx})_{00}$, and for $l \ne 0,~\tilde l\ne \tilde 0$ also
$\epsilon_{\tilde l},~
\epsilon_l-\epsilon_{\tilde 0}
\to \epsilon_l$ as well as  $(j^{-q}_x)_{\tilde 0l}, (j^{-q}_x)_{\tilde l 0}, 
(j^{0}_x)_{\tilde 0 \tilde l} \to (j^0_{x})_{0l}$.
Comparing Eqs.(\ref{eq8},\ref{eq10}) we recognize
\begin{equation}
\gamma_{\tilde 0} = 2 (j^q_y)_{0\tilde 0} D_{xx}. \label{eq11}
\end{equation}

There is no analogous simple $q \to 0$ limit for $(j^q_y)_{0\tilde
0}$.  Useful relation is obtained when we consider for $B=0$ the
current response $j^q_y$ to external field ${\cal E}^q_y$. Such a
response has been invoked for finite system in order to enforce the
validity of the sum rule for $\sigma^{reg}_{\alpha\alpha}$ \cite{loh}.
The corresponding conductivity $\sigma^{qq}_{yy}$ can be expressed in
analogy to the $q=0$ one in Eqs.(\ref{eq8}),
\begin{equation}
\sigma^{qq}_{yy}(\omega)= {i e^2 \over \Omega}
\sum_{\tilde m} {|(j_y^q)_{0\tilde m}| ^2 \over \epsilon_{\tilde m}}
[{1\over \omega^+ - \epsilon_{\tilde m}} +{1\over \omega^+ + 
\epsilon_{\tilde m}}] .\label{eq12}
\end{equation}
The essential difference to Eqs.(\ref{eq8}) is that the coherent peak
at $\omega=0$ now splits into two peaks at  $\omega=\pm
\epsilon_{\tilde 0}$, respectively. Since the sum rule is not 
changed \cite{loh}, we can equate their intensities \cite{eder},
\begin{equation}
D_{yy}   = (j^q_y)_{0\tilde 0}^2 /\epsilon_q. \label{eq13}
\end{equation}
Moreover, for a single QP also the coherent mass is directly related
to the stiffness \cite{zoto}, i.e. $\epsilon_q=D_{yy} q^2$. Hence we
get from Eq.(\ref{eq13}) $|(j^q_y)_{0\tilde 0}| = D_{yy} q$.  From
Eqs.(\ref{eq7},\ref{eq9}), with $K^{q}_{yx}(\omega > \epsilon_{\tilde
0}) \sim 2\gamma_{\tilde 0} /\omega^+$, we finally obtain
\begin{equation}
R_H^0 = {\rm sgn}(\zeta){\Omega \over e}, 
\quad \zeta=\langle 0|j^q_y|\tilde 0 \rangle.\label{eq14}
\end{equation}
This is the semiclassical result for a single QP, since 
$Z= {\rm sgn}(\zeta)=\pm 1$. 

The remaining question of the sign of $\zeta$ is not trivial, at least
we did not find a simple argument which would yield the expected
plausible answer.  While it is easy to show that $Z=1$ for a single
free electron at bottom of the band (note that $e=-e_0<0$), for more
general case $Z$ requires the knowledge of the ground state wavefunction
$|\Psi^q\rangle$ at finite $q \ne 0$, which can be quite involved
within a correlated system. For a single QP it is convenient to
represent $|\Psi^q\rangle$ in terms of localized functions
\begin{equation}
|\Psi^{q}\rangle = \sum_l {\rm e}^{i({\vec Q+\vec q})
\cdot \vec r_l} |\psi^q_l\rangle. \label{eq15}   
\end{equation}
Then one can express $\zeta$ via Eq.(\ref{eq3}) for $q\ll 1$, assuming
a local character of $|\psi^q_l\rangle$,
\begin{equation}
\zeta = {qt \over 2}\sum_{ijl} e^{- i\vec Q\cdot \vec r_l} \langle 
\psi^q_l| (r^y_i +r^y_j)(j^{ij}_{y+}-j^{ij}_{y-})
|\psi^q_0 \rangle,
\label{eq16}
\end{equation}
where we have chosen $\vec r_0=0$, and $j^{ij}_{y\pm} = 
\sum_s r^y_{ij} c^{\dagger}_{js} c_{is}$  are forward
(backward) hopping operators in Eq.(\ref{eq3}), corresponding to
$r^y_{ij}=\pm 1$, respectively.  It seems now plausible that the
charge character should come from the sign of the dominating forward
hopping $r^y_i+ r^y_j$ in Eq.(\ref{eq16}), which should be positive
for an electron (e.g. for a free electron the only forward term is
$\vec r_i=0, \vec r_j=\vec r_l = \vec e_y$), and negative for holes
due to the opposite direction of the electron hopping (at least for
free fermions). However as shown below for the concrete example the
evaluation and the result is not so evident in general.

Let us consider the problem of the spin polaron, i.e. a single hole
inserted into the AFM \cite{schm}, as relevant to cuprates and
described within the $t-J$ model,
\begin{equation}
H= - t\sum_{\langle ij\rangle s} (\tilde c^{\dagger}_{js} 
\tilde c_{is} + H.c.) +J \sum_{\langle ij\rangle } 
\vec S_i \cdot \vec S_j, \label{eq17}
\end{equation}
where $\tilde c_{is},\tilde c^{\dagger}_{is}$ are projected operators,
not allowing for the double occupancy of sites. Although the
projection introduces the interaction also in the kinetic term, the
analysis of the Hall response presented above remains valid, provided
that operators $\tilde j$ and $\tau$ in Eqs.(\ref{eq3}) are redefined
accordingly.

Quantity $\zeta$ for a single hole in the $t-J$ model has been
considered in another context in Ref.\cite{eder}.  Since in general
$|\psi^q_l\rangle$ are complicated, we evaluate $\zeta$ within the
perturbation expansion. We start with a static hole in the N\'eel AFM,
taking into account only the Ising-type spin interaction $J S_i^z
S_j^z$.  Corrections due to the hopping term $H_{kin}$ with $t/J \ll
1$ and due to the spin flip part $H_{\gamma}$ with $\gamma=J_{\perp}/J
\ll 1$ are treated at $T=0$ perturbatively \cite{prel}. As noted in
Ref.\cite{eder} nonzero contributions in Eq.(\ref{eq16}) can arise
only from nonlocal $l\ne 0$ terms, since for $l=0$ contributions of
$j_{y-}$ and $j_{y+}$ cancel. The ground state of a hole in the $t-J$
model is at finite $\vec Q=(\pm
\pi/2,\pm\pi/2)$ \cite{schm,prel}. Although such a ground state is
clearly degenerate, this should not change our results
for $R_H^0$.

It is convenient to represent localized wavefunctions
$|\psi^q_l\rangle$ in terms of basic (string) states
$|\varphi^q_{lm}\rangle$, which are obtained by applying on the ground
state $|\varphi^0_{l0}\rangle$ (N\'eel state with a static hole on the
site $l$) operations of $H_{kin}$ and $H_{\gamma}$. Finally we want to
express
\begin{equation}
|\psi^q_l\rangle =\sum_m c^q_m |\varphi^{q}_{lm}\rangle. \label{eq18} 
\end{equation}
Although the calculation of allowed $|\varphi^{q}_{lm}\rangle$ and the
corresponding $c^q_m$ can be quite involved (and not unique), it is
straightforward within lowest orders of the perturbation series in
$t,\gamma$ \cite{prel,eder}.  The lowest nonzero contribution in
Eq.(\ref{eq14}) comes e.g. from $\vec r_l=-2\vec e_y$. Let us evaluate
the term starting with $|\varphi^{0}_{00}\rangle$. The operator
$~j^+_{0j}$ with $\vec r_j=-\vec e_y$ moves the hole in the $-y$
direction and leaves behind one flipped spin (relative to the N\'eel
state).  The resulting state can be represented also as an excited
state $|\varphi^{q}_{lm}\rangle$, reached from
$|\varphi^{0}_{l0}\rangle$ within the perturbation expansion by a
successive application of a single $H_{kin}$ and $H_{\gamma}$. This
particular contribution within the perturbation expansion is thus
\begin{equation}
\zeta_{ijl}= {\rm e}^{-i\vec Q \cdot \vec r_l}{qt^2\over 2\Delta_1
\Delta_2} < 0, \label{eq19}
\end{equation}
where $\Delta_1,\Delta_2>0$ are energies of intermediate excited
states. There are more nonzero contributions within the same order of
the perturbation theory, but they are as well negative, confirming the
hole-like character of the QP. Although the result $Z=-1$ is strictly
valid in the $t-J$ model for $\gamma \ll 1,~J/t\ll 1$, we do not
expect any change of the sign entering the relevant regime $\gamma=1,
~J<t$
\cite{prel}.

Obtained results are not surprising. The charge carrier in an
interacting system, obeying the properties of the QP, behaves in the
d.c. Hall response at $T=0$ according to the semiclassical
relation. The main novelty is that no particular assumptions as the
relaxation time approximation are needed to derive this result.  This
should hold for correlated systems, but as well as to electron
interacting with phonons etc. We presented also the calculation which
confirms the hole-like Hall response for a single hole in the $t-J$
model.

It is tempting to generalize above results in several directions, in
order to be applicable to more realistic situation in correlated
systems, and in cuprates in particular. At low hole doping $n_h\ll 1$
we expect that holes in cuprates behave as independent QP - spin
polarons. Then $\tilde \sigma_{\alpha\beta} \propto n_h$, which leads
to $R_H^0 = 1/n_h e_0$ in this regime. Such a behavior seems to be
indeed found at lowest $T$ in the underdoped $La_{2-x}Sr_xCuO_4$
\cite{ong,hwan}  (although there seem to be quantitative 
discrepancies between various data), but not quite so in all other
underdoped cuprates \cite{ong}.  Approaching the `optimum'-doping
regime the scenario of independent QP is clearly not applicable, since
electrons seem to reveal a rather well defined large Fermi surface,
while $R_H^0$ can even change sign. The evaluation of $R_H^0$ in this
intermediate regime can be in principle treated with
Eqs.(\ref{eq7}-\ref{eq10}), taking into account in Eq.(\ref{eq9}) all
relevant low lying excited states $|\tilde m\rangle$.  Whereas the
stiffness $D_{\alpha\alpha}$ has been considered quite in detail
before \cite{scal}, looking to Eq.(\ref{eq11}) the central quantity
for $R_H^0$ appears to be $\langle (j^q_y)_{0\tilde m}\rangle_{av}$,
averaged over low-lying excitations satifying $\epsilon_{\tilde
m}(q\to 0) \to 0$. Analysing the latter quantity in the doped system
one could possibly gain more insight into the change of
the charge-carrier character on doping, so far theoretically
understood only for the high-frequency quantity $R_H^*= R_H(\omega \to
\infty)$ \cite{shas,assa}. Another challenging question is clearly the
anomalous $R_H^0(T)$ dependence, which is however beyond our $T=0$
analysis.
 
The author acknowledges the financial support and the hospitality of
the Max-Planck Institut f\"ur Festk\"orperforschung, Stuttgart, as well as
of the Institut Romand des Recherches Num\'erique en Physiques des
Mat\'eriaux (IRRMA), EPFL, Lausanne, where part of this work has been
performed.  The author also thanks P. Horsch, A. Ram\v sak and X. Zotos
for fruitful discussions.


\begin{references}
\bibitem{ong} For a review see e.g. N. P. Ong, in {\it Physical
Properties of High Temperature Superconductors}, ed. by D. M. Ginsberg
(World Scientific, Singapore, 1990), Vol. 2, p.459.
\bibitem{carr} A. Carrington, A. P. Mackenzie, C. T. Lin, and
J. R. Cooper, Phys. Rev. Lett. {\bf 69}, 2855 (1992).
\bibitem{hwan} H. Y. Hwang, {\it et al.}, Phys. Rev. Lett. 
{\bf 72}, 2636 (1994).
\bibitem{hols} T. Holstein and L. Friedman, Phys. Rev.
{\bf 165}, 1019 (1968).
\bibitem{brin} W. F. Brinkman and T. M. Rice, Phys. Rev. B
{\bf 4}, 1566 (1971).
\bibitem{fuku} H. Fukuyama, H. Ebisawa, and Y. Wada, Prog.
Teor. Phys. {\bf 42}, 494 (1969).
\bibitem{itoh} M. Itoh, J. Phys. F {\bf 14}, L89 (1984);
{\bf 15}, 1715 (1985); H. Kohno and K. Yamada, Prog. Teor. Phys.
{\bf 80}, 623 (1988).
\bibitem{ioff} L. B. Ioffe, V. Kalmeyer and P. B. Wiegmann, 
Phys. Rev. B {\bf 43}, 1219 (1991); C. A. R. Sa de Melo, 
Z. Wang, and S. Doniach,
Phys. Rev. Lett. {\bf 68}, 2078 (1992).
\bibitem{shas} B. S. Shastry, B. I. Shraiman, and R. R. P. Singh,
Phys. Rev. Lett. {\bf 70}, 2004 (1993).
\bibitem{cast} H. E. Castillo and C. A. Balseiro, Phys. Rev. Lett.
{\bf 68}, 121 (1992).
\bibitem{assa} F.F. Assaad and M. Imada, Phys. Rev. Lett. {\bf
74}, 3868 (1995).
\bibitem{kohn} W. Kohn, Phys. Rev. {\bf 133}, A171 (1964).
\bibitem{scal} B. S. Shastry and B. Sutherland,
Phys. Rev. Lett. {\bf 65}, 243 (1990); D. J. Scalapino, S. R. White
and S. Zhang, Phys. Rev. B {\bf 47}, 7995 (1993).
\bibitem{schm} S. Schmitt-Rink, C, Varma, and A. Ruckenstein,
Phys. Rev. Lett. {\bf 60}, 2783 (1988); C. L. Kane, P. A. Lee,
and N. Read, Phys. Rev. B {\bf 39}, 6880 (1989);
G. Martinez and P. Horsch, Phys. Rev. B {\bf 44}, 317 (1991).
\bibitem{prel} P. Prelov\v sek, I. Sega, and J. Bon\v ca,
Phys. Rev. B {\bf 42}, 10706 (1990).
\bibitem{eder} R. Eder, Phys. Rev. B {\bf 44}, 12609 (1991).
\bibitem{zoto} X. Zotos, P. Prelov\v sek and I. Sega, Phys. Rev.
B {\bf 42}, 8445 (1990).
\bibitem{loh} E. Y. Loh, T. Martin, P. Prelov\v sek and
D. K. Campbell, Phys. Rev. B {\bf 38}, 2494 (1988); I. Sega and
P. Prelov\v sek, Phys. Rev. B {\bf 42}, 892 (1990).
\end{references}
\end{document}